\documentclass[%
 reprint,
superscriptaddress,
 amsmath,amssymb,
 aps,
pra,
prb,
rmp,
prstab,
prstper,
floatfix,
]{revtex4-2}

\usepackage{caption}
\usepackage{color}
\usepackage{graphicx}
\usepackage{dcolumn}
\usepackage{bm}
\usepackage{subfigure}
\usepackage{float}

\begin{document}

\preprint{APS/123-QED}

\title{A Compact One-Way Fault-Tolerant Optical Quantum Computation}

\author{Peilin Du}
\affiliation{
State Key Laboratory of Quantum Optics Technologies and Devices, Shanxi University, Taiyuan, 030006, China}%
\affiliation{
College of Physics and Electronic Engineering, Shanxi University,Taiyuan, 030006, China}

\author{Jing Zhang}
\email{zjj@sxu.edu.cn}
\affiliation{
State Key Laboratory of Quantum Optics Technologies and Devices, Shanxi University, Taiyuan, 030006, China
}%
\affiliation{
College of Physics and Electronic Engineering, Shanxi University,Taiyuan, 030006, China
}%

\affiliation{
Collaborative Innovation Center of Extreme Optics, Shanxi University, Taiyuan, 030006, China
}%

\author{Tiancai Zhang}%
\affiliation{
State Key Laboratory of Quantum Optics Technologies and Devices, Shanxi University, Taiyuan, 030006, China
}%

\affiliation{
Collaborative Innovation Center of Extreme Optics, Shanxi University, Taiyuan, 030006, China
}%

\author{Rongguo Yang}%
\email{yrg@sxu.edu.cn}
\affiliation{
State Key Laboratory of Quantum Optics Technologies and Devices, Shanxi University, Taiyuan, 030006, China
}%
\affiliation{
College of Physics and Electronic Engineering, Shanxi University,Taiyuan, 030006, China
}%

\affiliation{
Collaborative Innovation Center of Extreme Optics, Shanxi University, Taiyuan, 030006, China
}%
\author{Jiangrui Gao}%
\affiliation{
State Key Laboratory of Quantum Optics Technologies and Devices, Shanxi University, Taiyuan, 030006, China
}%

\affiliation{
Collaborative Innovation Center of Extreme Optics, Shanxi University, Taiyuan, 030006, China
}%
             
\begin{abstract}
One-way quantum computation is a promising approach to achieving universal, scalable, and fault-tolerant quantum computation. However, a main challenge lies in the creation of universal, scalable three-dimensional cluster states. Here, an experimental scheme is proposed for building large-scale canonical three-dimensional cubic cluster states, which are compatible with the majority of qubit error-correcting codes, using the spatiospectral modes of an optical parametric oscillator. Combining with Gottesman-Kitaev-Preskill states, one-way fault-tolerant optical quantum computation can be achieved with a lower fault-tolerant squeezing threshold. Our scheme drastically simplifies experimental configurations, paving the way for compact realizations of one-way fault-tolerant optical quantum computation.
\end{abstract}

\maketitle


\section{\label{sec:introduction}introduction}

Continuous-variable (CV) quantum optical platform is a viable candidate for quantum computation\cite{Takeda(2019),Pfister(2019),Fukui(2021),Furusawa2024review}, as it allows for the deterministic generation of cluster states. Over the past decade, large-scale one-dimensional (1D) cluster states have been experimentally generated by frequency\cite{Pysher(2011),Chen(2014)} and time\cite{yokoyama(2013),yoshikawa(2016)} multiplexing, single-mode Gaussian operations and further scalable quantum operations were demonstrated\cite{Asavanant(2020)One-hundred}. Different kinds of large-scale two-dimensional (2D) cluster states have been generated in the experiment by time multiplexing\cite{asavanant(2019),larsen(2019)}, and a small quantum circuit consisting of 10 single-mode and 2 two-mode gates was executed on a three-mode input state\cite{Larsen(2020)quantum-computing}. Moreover, numerous theoretical schemes for generating large-scale three-dimensional (3D) cluster states have been proposed\cite{Fukui(2018)SURFACECODE,wu(2020),Fukui3D(2020),yangrg(2020),Bour2021blue,Zhuxuanoptical(2021),TzitrinPRXQuantum(2021),Larsen(2021)PRXQuantum,Du(2022)}. 
 
In the overwhelming majority of the aforementioned work, large-scale 2D cluster states are generated using a fixed, constant-depth circuit to interfere with two to four squeezed quantum combs, while the generation of large-scale 3D cluster states requires more sophisticated circuits. The resulting cluster states, referred to as macronode cluster states\cite{MenicucciTemporal(2011),PeiWeaving(2014)}, must be reduced to canonical cluster states for quantum computation implementation. Additionally, injecting the required non-Gaussian Gottesman-Kitaev-Preskill (GKP) states into cluster states relies on optical switches, but realizing practical optical switches for quantum computation remains challenging. These factors inherently increase the overall difficulty of achieving fault-tolerant quantum computation on a CV quantum optical platform. 

In this paper, we propose a compact scheme for directly generating large-scale canonical 1D linear and 2D planar cluster states from an optical parametric oscillator (OPO), which are key resources for quantum computation. Large-scale canonical 3D cubic cluster states, which allow topological quantum error correction encoding, can be further generated by using phase modulation (PM)\cite{Zhuxuanoptical(2021)}. Our scheme is compatible with the Photon-counting-Assisted Node-Teleportation Method (PhANTM) protocol, allowing GKP states to be reliably generated and embedded within cluster states through the addition of local photon-number-resolving (PNR) measurements\cite{Miller2022GKP}. By combining the non-Gaussian GKP states with the Raussendorf-Harrington-Goyal (RHG) code\cite{RAUSSENDORF(2006),RaussendorfPRL(2007),Raussendorf(2007)}, one-way fault-tolerant optical quantum computation can be achieved, with a corresponding fault-tolerant squeezing threshold of 9.4 dB. Our scheme drastically simplify experimental configurations, paving the way for compact realizations of one-way fault-tolerant optical quantum computation on CV quantum optical platform. Additionally, the compactness of our scheme offers a viable option for implementing one-way fault-tolerant quantum computation on a chip using quantum nanophotonics\cite{VaidyaCHIP,DioumCHIP24}.

The paper is organized as follows. Section II describes the generation of large-scale canonical 1D linear, 2D planar, and 3D cubic cluster states from an OPO. Section III discusses topological fault-tolerant quantum computation using the RHG-GKP code. The main results are summarized and further discussed in Section V.

\section{Generation of large-scale canonical 1D linear, 2D plane and 3D cubic cluster states}

In our scheme, OPO is a self-imaging or large-Fresnel-number cavity with a type II phase-matching nonlinear crystal (second-order nonlinear coefficient is $\xi$) in it, which can be simultaneously resonant for many transverse modes, and its transverse eigenmodes are the complete set of Hermite-Gaussian (HG) modes; longitudinal eigenmodes are spaced by the free spectral range (FSR)\cite{largeFresnelnumber(2009)}. This specially designed OPO can guarantee simultaneous and sustainable nonlinear interaction and resonance of all down-converted optical frequency comb modes\cite{Multimoderesonance(2010)}. Furthermore, sideband modes are generated around each comb mode of the optical frequency comb by the OPO\cite{BarbosaPRLsideband(2018)}. 

\begin{figure}[t]
   \centering
    {\includegraphics[width=0.9\linewidth]{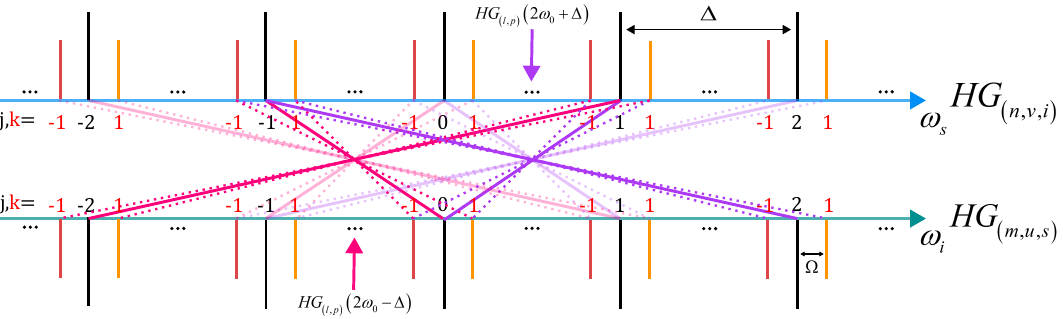}}
    \caption{Schematic for generating 1D linear H graph states.}
    \label{fig5}
\end{figure}

As shown in Fig.\ref{fig5}, two spatially structured pump fields HG$_{\left(l,p\right)}$ with frequencies $\omega_{b}=2\omega_{0}\pm \Delta$ (purple and pink arrows) can be down-converted into signal HG$_{\left(m,u,s\right)}$ (blue line) and idler HG$_{\left(n,v,i\right)}$ (green line) optical frequency combs. The signal and idler comb modes (black lines), along with their sideband modes (yellow and red lines), have frequencies $\omega_{s}=\omega_{i}=\omega_{0}\pm\left(j\Delta+k\Omega\right)$, $j$ and $k\in\left\{0, 1, 2, \dots \right\}$. $\Delta$ represents the FSR of the OPO on the order of THz, while $\Omega$ denotes the frequency spacing between the sideband modes on the order of MHz, the first subscripts $\left\{l,m,n\right\}$ and the second subscripts $\left\{p,u,v\right\}$ represent azimuthal index (an integer) and the radial index (a non-negative integer), respectively. The nonlinear interaction must satisfy the conservation of energy $\left(\omega_{b}=\omega_{s}+\omega_{i}\right)$, momentum $\left(\vec{\mu}_{b} =\vec{\mu}_{s}+\vec{\mu}_{i}\right)$, and order number ($l+p=m+u=n+v$). 

The Hamiltonian of this nonlinear interaction under the undepleted pump approximation is the following:
\begin{equation}
\mathrm{H}=i\hbar\xi\sum_{j,k} \mathrm{G}\left(a_{\left(m,u,s\right)}a_{\left(n,v,i\right)}-\mathrm{H}.c.\right),
\label{eq1}
\end{equation}
where $a_{\left(m,u,s\right)}$ is the annihilation operator of down-converted signal field HG$_{\left(m,u,s\right)}$ with frequency $\omega_{s}$. The $\mathrm{G}$ matrix specifies the Hamiltonian for the OPO that acts on the vacuum of the cavity modes to generate the Hamiltonian graph ($\mathrm{H}$ graph) state and denotes the adjacency matrix of this state\cite{Menicucci2007star,Menicucci2011star,menicucci(2011)Hgraph}. It depends on the modal content of the pump field and on the overlap integral of the coupling among pump, signal, and idler fields\cite{AlvesRulePRA}. Each pair of comb (sideband) modes connected by a purple or blue solid (dashed) line are from the same nonlinear interaction process as shown in Fig.\ref{fig5}. These connections can simultaneously form two 1D linear H graph states (depicted as dark and light lines) for different $\left|k\right|$. Note that these H graph states resulting from Eq.(\ref{eq1}) can always be representable as cluster states\cite{menicucciAmatrix}. 

To gain perspective on the overall structure of generated cluster states, we streamline the analysis by focusing on a single linear H graph state comprising 48 modes. According to the interaction Hamiltonian of Eq.(\ref{eq1}), the G matrix can be written as shown in Fig.\ref{fig2}(a). The adjacent matrix (A matrix) of generated cluster state is obtained using the method outlined in Ref.\cite{menicucciAmatrix} (more details are shown in Appendix A), and displayed in Fig.\ref{fig2}(b). The generated cluster state is nominally a complete bicolorable state (i.e., modes 1-24 are not linked to one another but are all linked to all of the modes 25-48). We further reduce A matrix after rounding down all elements below a certain weight to zero, the weights being chosen to correspond to realistic values of squeezing: Fig.\ref{fig2}(c) displays the results of such graph pruning for the weight range 0.5 to 0.35 (squeezing range -3 to -4.5 dB)\cite{yangrg(2020)}. In the figure, blue (green) circles represent down-converted signal (idler) fields. The resulting cluster state is a canonical 1D linear cluster state. Remarkably, even lower squeezing amounts are sufficient to generate this canonical 1D linear cluster state, which serves as a universal structure for single-mode quantum computation.
\begin{figure}[t]
   \centering
    {\includegraphics[width=0.8\linewidth]{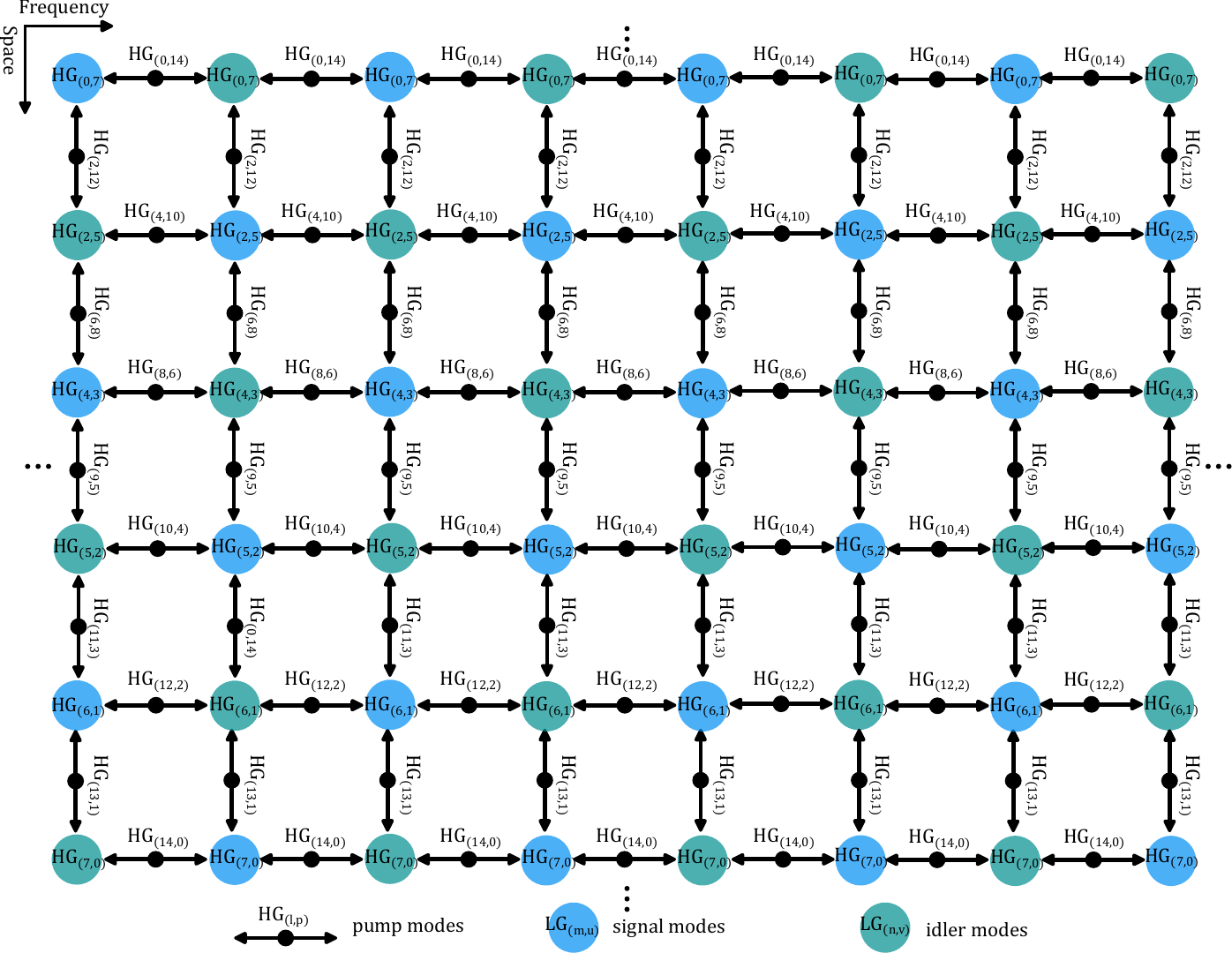}}
    \caption{The concurrent down-conversion processes are allowed for the complex spatial HG mode of Eq.(\ref{eq2}).}
    \label{fig1}
\end{figure} 

In the generation process of the canonical 1D cluster states discussed above, the pump fields possess only a single spatial structure. For the purpose of generating large-scale canonical 2D cluster states, the OPO is pumped by two complex spatially structured fields HG$_{\left(l,p\right)}$ with frequencies $2\left(\omega_{0}\pm \Delta\right)$, whose electric field can be written in the HG basis as,
\begin{eqnarray}
&\mathrm{HG}_{\left(l,p\right)}=a*\mathrm{HG}_{\left(0,14\right)}+b*\mathrm{HG}_{\left(2,12\right)}+c*\mathrm{HG}_{\left(4,10\right)} \nonumber \\ &+d*\mathrm{HG}_{\left(6,8\right)}
+e*\mathrm{HG}_{\left(8,6\right)}+f*\mathrm{HG}_{\left(9,5\right)}+\dots,
\label{eq2}
\end{eqnarray}
where $\left\{a, b, c,\dots \right\}$ are the amplitude coefficients. Similarly to the 1D case, these down-conversion processes can simultaneously form two 2D planer H graph states for different $\left|k\right|$. Here, we consider only a 2D planar H graph state with 48 modes, as shown in Fig.\ref{fig1}. The blue (green) circles represent down-converted signal (idler) fields, while the black double-sided arrows represent pump fields. For example, the modal content HG$_{\left(0,14\right)}$ of the pump field has down-conversion process: $\mathrm{HG}_{\left(0,14\right)}\to \mathrm{HG}_{\left(0,7,s\right)}+\mathrm{HG}_{\left(0,7,i\right)}$. By appropriately tuning the amplitude coefficients in Eq.(\ref{eq2}), these concurrent down-conversion processes can be ensured to have the same intensity.  
\begin{figure*}[htbp]
\centering\includegraphics[height= 0.45\textheight, width=0.8\linewidth]{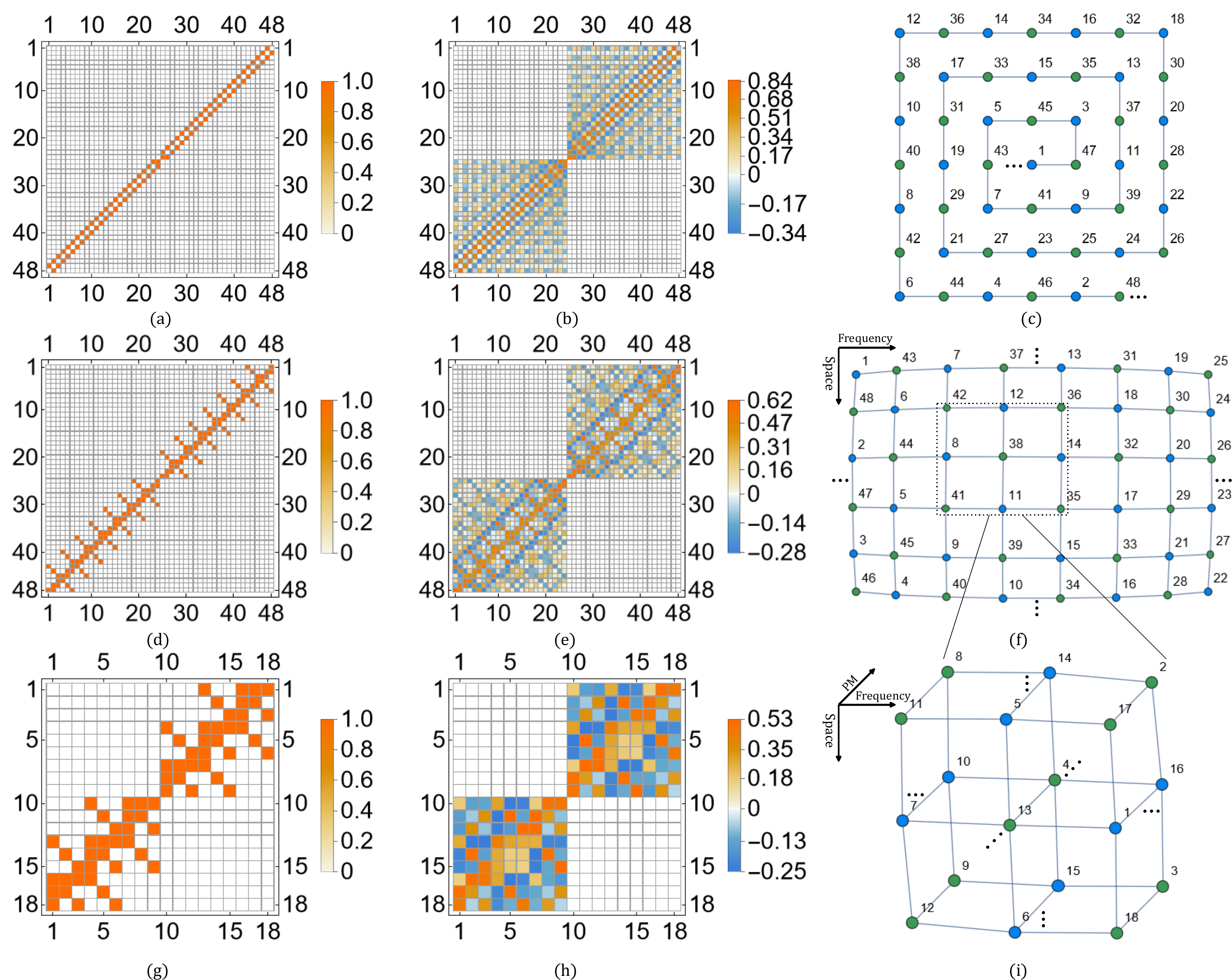}
\caption{The G matrix of the 1D linear (a), 2D planar (d) and 3D cubic (g) H graph states, the corresponding A matrix ((b), (e) and (h)) for generating cluster states. The resulting canonical 1D linear (c), 2D planar (f) and 3D cubic (i) cluster states after reducing A matrix. The blue (green) circles represent down-converted signal (idler) fields.}
\label{fig2}
\end{figure*}

Based on the concurrent down-conversion processes depicted in Fig.\ref{fig1}, the G matrix can be expressed as shown in Fig.\ref{fig2}(d). The generated cluster state also is a complete bicolorable state, and its A matrix is shown in Fig.\ref{fig2}(e). Fig.\ref{fig2}(f) illustrates the cluster state obtained by reducing the A matrix within the weight range of 0.31 to 0.29 (corresponding to a squeezing range of -5.1 to -5.4 dB). The resulting cluster state is a canonical 2D planar cluster state with each mode corresponding to the mode in Fig.\ref{fig1}, which is a universal resource for quantum computation\cite{VanUniversalResources}.

In order to generate 3D cluster states, the 2D planar H graph states are entangled with neighboring 2D planar H graph states using PM, which can be achieved using either an intrinsically phase-modulated OPO or an externally modulated OPO. An effective Hamiltonian of PM can be written as\cite{Zhuxuanoptical(2021)}:
\begin{equation}
    \mathrm{H}=i\hbar\chi\sum_{j,k}  \mathrm{G}\left(a_{k}a_{k+1}^{\dagger}-\mathrm{H}.c.\right), 
    \label{eq3}
\end{equation}
where $\chi$ is related to the modulation index, phase and interaction time of PM, $a_{k}$ is the annihilation operator of down-converted field with frequency $\omega_{0}\pm\left(j\Delta+k\Omega\right)$, while $a^{\dagger}_{k+1}$ is the creation operator of down-converted field with frequency $\omega_{0}\pm\left(j\Delta+\left(k+1\right)\Omega\right)$. After the PM, all 2D planer H graph states are connected to form two independent 3D cubic H graph states. Here, a minimum unit (18 modes) of a 3D cubic H graph state is considered for simplicity. By cascading the two different quantum evolutions Eq.(\ref{eq1}) and Eq.(\ref{eq3}), the elements of G matrix can be further increased and written as shown in Fig.\ref{fig2}(g), the corresponding A matrix of generated cluster state is displayed in Fig.\ref{fig2}(h). As shown in Fig.\ref{fig2}(i), the resulting cluster state is a canonical 3D cubic cluster state with a weight of approximately 0.35 (squeezing around -4.5 dB), which is compatible with various qubit error-correcting codes, including the color, surface, hyperbolic surface, and RHG codes. 

Because the spatial and spectral properties of eigenmodes are reasonably independent, and the Gouy phase is identical for spatial modes of the same order in the absence of dispersion, we consider all seventh-order spatial modes produced by down-conversion during the generation of canonical 2D planar and 3D cubic cluster states that can resonate within a single cavity. The maximum number of canonical 1D linear, 2D planar, and 3D cubic cluster states that can be generated in parallel depends on the number of sideband modes, which, in turn, is determined by the ratio of $\Delta$ to $\Omega$, allowing for up to $10^{3}$ modes. The scale of generated canonical 2D planar and 3D cubic cluster states can be significantly expanded by increasing the spatial and spectral modal content. The maximum number of spectral modes is determined by the ratio of the phase-matching spectral bandwidth with $\Delta$, potentially reaching $10^{3}$ modes\cite{PWang(2014)}. The maximum number of spatial modes is constrained by the experimentally available HG modes, which can also yield $10^{1}$ modes\cite{Higher-orderHermite-Gauss,High-order2}. Consequently, the mode numbers of the generated canonical 2D planar and 3D cubic cluster states in our scheme can reach $10^{7}$.

\section{Fault-tolerant quantum computation}

After generating the canonical 1D linear, 2D planar and 3D cubic cluster states, quantum computation can proceed as usual. However, non-Gaussian GKP states are required for universal quantum computation and quantum error correction. In other works, non-Gaussian GKP states are injected into cluster states using optical switches, which present significant experimental challenges. These switches must exhibit extremely low loss, minimal switching time, high repetition rates, and minimal optical nonlinearity. Our scheme is compatible with PhANTM protocol proposed by Miller\cite{Miller2022GKP}, where GKP states can be reliably generated using local PNR measurements. Nevertheless, the GKP code can only correct the small displacement errors (usually less than $\sqrt{\pi}/2$) in phase space, while the large displacement errors are converted into Pauli errors in the encoded qubits\cite{GKPPRA,KyungjooPRL(2020)}. Furthermore, residual errors can be further treated by combining the GKP code with certain higher-level qubit error-correcting codes, such as color code\cite{ChamberlandColorcode2020,ZhangPRAcolorcode}, surface code\cite{Fukui(2018)SURFACECODE,Vuillot(2019)SURFACE,Noh(2019)SURFACE}, hyperbolic surface code\cite{BreuckHyperbolicSurfaceCodes} or RHG code\cite{RAUSSENDORF(2006),RaussendorfPRL(2007),Raussendorf(2007)} towards fault-tolerant MBQC.
\begin{figure}[h]
\centering\includegraphics[width=\linewidth]{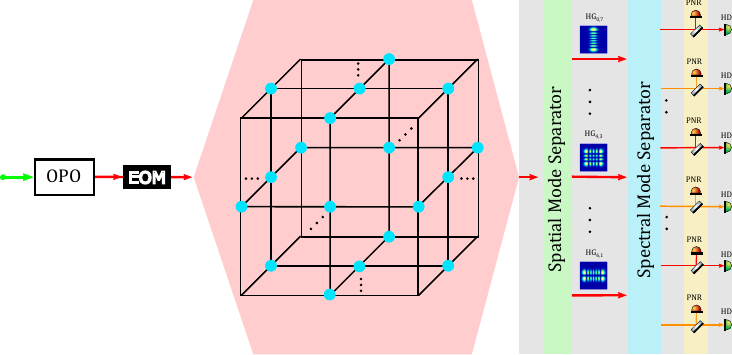}
\caption{The setup of the compact one-way fault-tolerant optical quantum computation on CV quantum optical platform.}
\label{fig3}
\end{figure}

Due to its high-noise threshold for fault-tolerant quantum computation, we focus on concatenation the GKP code with the RHG code to achieve topological fault-tolerant quantum computation, although our scheme can readily accommodate other error-correcting codes. As shown in Fig.\ref{fig3}, we propose an architecture for compact one-way fault-tolerant quantum computation on CV quantum optical platform. Leveraging the experimental simplicity of placing an electro-optic modulator (EOM) after an OPO, we employ an externally modulated OPO to generate canonical 3D cubic cluster states. Spatial and spectral mode separators are then utilized to solely separate the modes of the canonical 3D cubic cluster states for measurement. The spatial mode separator (green region) extracts the spatial HG modes, while the spectral mode separator (blue region) further isolates the spectral modes. By cascading these two separators, each mode of the 3D cubic cluster states can be transmitted through a single beam. Subsequently, local PNR measurements in the yellow region are employed to embed GKP states into the 3D cubic cluster states via the PhANTM protocol. Finally, fault-tolerant quantum computation is performed through homodyne detection (HD) of all modes.

To establish a fault-tolerant threshold, we numerically simulate the complete architecture. Finite squeezing noise is modeled using a Gaussian random displacement channel applied to the cluster and GKP states\cite{GKPPRA},
\begin{equation} N\left[\sigma\right]\left(\hat{\rho}\right)=\int\frac{d^{2}\alpha}{\pi\sigma^{2}}\mathrm{exp}\left[-\frac{\left|\alpha\right|}{\sigma^{2}}\right]\hat{D}\left(\alpha\right)\hat{\rho}\hat{D}^{\dagger}\left(\alpha\right), 
\end{equation}
where $\hat{D}\left(\alpha\right)$ is the displacement operator, and $\sigma^{2}$ represents the variance of the random displacement. Similarly, the loss noise introduced by the spatial and spectral mode separators is modeled as a Gaussian random displacement channel with variance $\sigma^{2}_{loss}=\left(1-\eta\right)/2\eta$, where $\eta$ is the total transmission coefficient of accumulated losses acting before each HD\cite{FukuiPRRLOSS}. 

In our architecture, finite squeezing noise is modeled as a Gaussian random displacement with variance $\sigma^{2}_{fin}=e^{-2r}/2$, while loss noise is characterized by a variance $\sigma^{2}_{loss}$ with $\eta=0.9$. The combined effects of finite squeezing and loss noise yield a total variance given by $\sigma^{2}_{total}=\sigma^{2}_{fin}+\sigma^{2}_{loss}$. Following the approach in Ref.\cite{TzitrinPRXQuantum(2021)}, we employ a maximum likelihood decoding strategy during the GKP error correction step and use the minimum weight perfect matching decoder to decode the RHG code, incorporating analog information from the GKP quantum error correction. Finally, logical qubit error rates are simulated as a function of squeezing using the Monte Carlo method with up to 100,000 samples.  
\begin{figure}[]
   \centering{\includegraphics[width=0.9\linewidth]{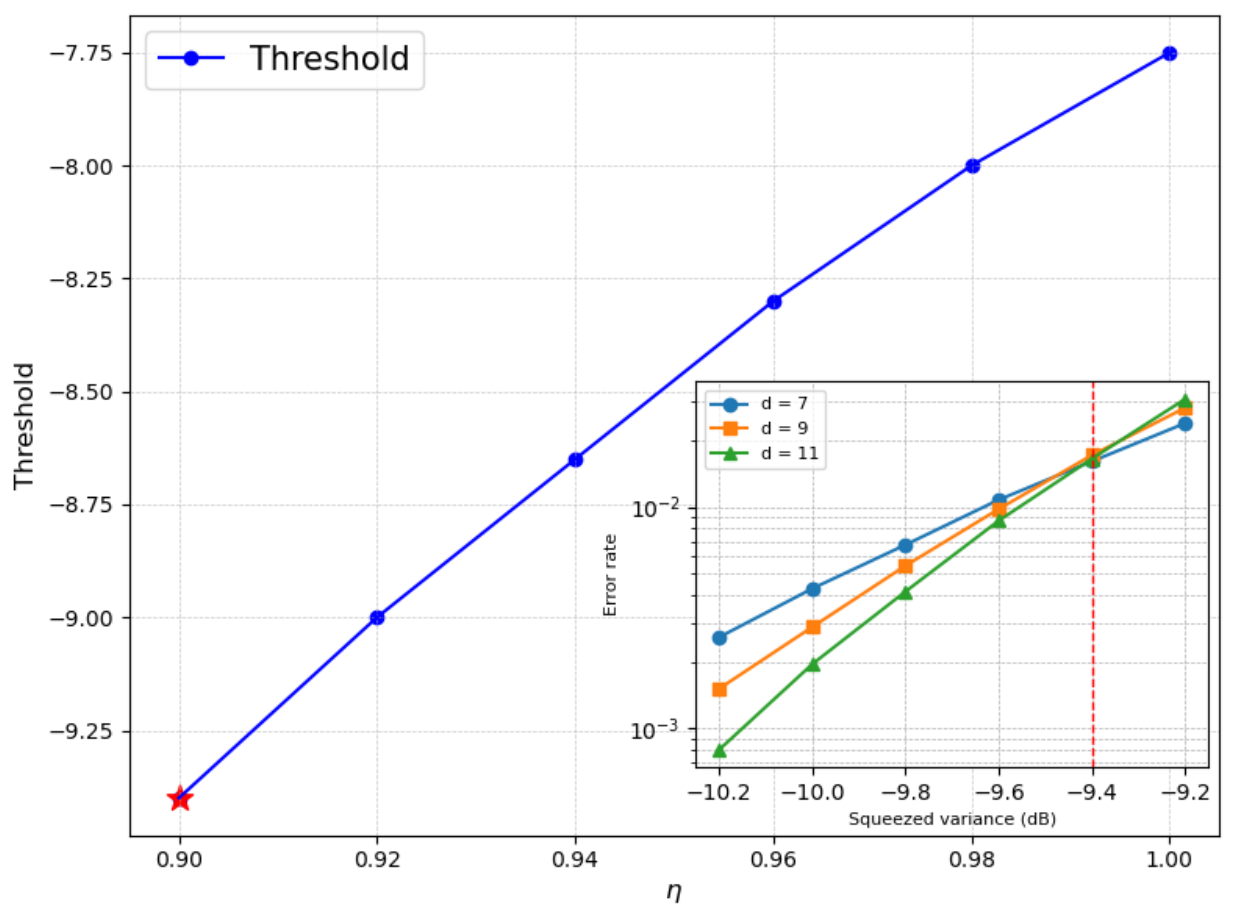}}
    \caption{The squeezing threshold of the RHG-GKP code as a function of the parameter $\eta$. Inset: when $\eta=0.9$, the error probability over squeezing for the RHG-GKP code with the distances $d\in\left\{7, 9, 11\right\}$, and the fault-tolerant squeezing threshold is 9.4 dB.}
    \label{fig4}
\end{figure}

As shown in Fig.\ref{fig4}, we present the squeezing threshold of the RHG-GKP code vary with the parameter $\eta$. The effect of total transmission coefficient $\eta$ on the threshold is monotonic: when loss noise is negligible ($\eta=1$), the squeezing threshold can reach 7.75 dB. For $\eta=0.9$ (inset of Fig.\ref{fig4}), the fault-tolerant squeezing threshold is 9.4 dB. In comparison, the squeezing thresholds for fault-tolerant quantum computation using the RHG-GKP code reported in Ref.\cite{Bour2021blue} and Ref.\cite{TzitrinPRXQuantum(2021)} are 10.5 dB and 10.1 dB, respectively. Our lower squeezing threshold reflects a significant improvement, which we attribute to the high performance of the generated 3D cubic cluster states. Moreover, the squeezing threshold can be further reduced by systematically considering the effects of impedance matching, linewidth, and sideband suppression ratio of the spectral mode separator on the modes when designing and constructing the spectral mode separator, thereby minimizing loss noise in our architecture\cite{ShiPRL2020}.

\section{Conclusion}
In this work, we propose a compact one-way optical quantum computation, which represents a promising candidate for universal, scalable, and fault-tolerant quantum computation. In our architecture, canonical 1D linear and 2D planar cluster states can be parallelly generated using only a single OPO, and canonical 3D cubic cluster states can be further constructed by cascading an external EOM. The 2D planar cluster states serve as universal resources for quantum computation, while the 3D cubic cluster states can readily accommodate the majority of error-correcting codes. By encoding the RHG-GKP code, we show that the squeezing threshold of fault-tolerant quantum computation is 9.4 dB and has room for improvement, taking into account finite squeezing and loss noise. This work provides a possible thought for realizing a compact fault-tolerant one-way quantum computation in experiments.

\section*{acknowledgments}
This work is supported by the Postgraduate Education Innovation Program of Shanxi Province (Grants No. 2024KY018), the National Key Research and Development Program of China (Grants No. 2021YFC2201802,  2021YFA1402002); the National Natural Science Foundation of China (Grants No. 11974225, No. 11874248 and No. 11874249), and the National Key Laboratory of Radar Signal Processing (Grant No. JKW202401).

\appendix
\section{Construct A matrix from G matrix}

Here, we detail the procedure for obtaining the A matrix from the G matrix, as described in Refs.\cite{menicucciAmatrix}. The process begins by diagonalizing the G matrix:
\begin{equation}
    \mathrm{G}=\mathrm{VDV^{T}},
\end{equation}
where the D is a diagonal matrix composed of the eigenvalues $\left\{\lambda_{1},\dots,\lambda_{2n}\right\}$ of the G matrix, while the V is constructed from the eigenvectors $\left\{\upsilon_{1},\dots,\upsilon_{2n}\right\}$ of the G matrix. The matrices D and V can be expressed as:
\begin{equation}
    \mathrm{D}=\begin{pmatrix}
\lambda_{1}&0  &0 \\
 0 & \cdots  & 0\\
  0& 0 &\lambda_{2n}
\end{pmatrix},\quad  \mathrm{V}=\begin{pmatrix}
\upsilon_{1},\dots,\upsilon_{2n} 
\end{pmatrix}.
\end{equation}
Then, the matrix V is separated into block matrix:
\begin{equation}
    \mathrm{V}=\begin{pmatrix}
 \mathrm{V}_{11} & \mathrm{V}_{12} \\
 \mathrm{V}_{21} & \mathrm{V}_{22}
\end{pmatrix}.
\end{equation}
The corresponding relation between the elements of A and G matrix is:
\begin{equation}
    \mathrm{A}_{0}=-\mathrm{V}_{12}\left(\mathrm{V}_{22}\right)^{-1}.
\end{equation}
Finally, we wind up the A matrix with
\begin{eqnarray}
  \mathrm{A}=&&\begin{pmatrix}
 0 & \mathrm{A}_{0}\nonumber \\
\mathrm{A}_{0}^{\mathrm{T}} & 0
\end{pmatrix}\nonumber\\
=&&\begin{pmatrix}
 0 & -\mathrm{V}_{12}\left(\mathrm{V}_{22}\right)^{-1}\\
 -\left(\mathrm{V}_{22}^{-\mathrm{T}}\mathrm{V}_{12}^{\mathrm{T}}\right) & 0
\end{pmatrix}.  
\end{eqnarray}

\nocite{*}
\bibliography{main}

\end{document}